\documentclass[a4paper,fleqn,usenatbib]{mnras}

\usepackage{mathptmx}

\usepackage[T1]{fontenc}
\usepackage{ae,aecompl}

\usepackage{graphicx}   
\usepackage{amsmath}    
\usepackage{amssymb}    

\newcommand{\Msun}{\ensuremath{{\rm M}_{\odot}}}

\newcommand{\avg}[1]{\ensuremath{\left\langle \,#1\, \right\rangle}}

\newcommand{\der}{\ensuremath{{\rm d}}}

\newcommand{\eqn}[1]{equation~\eqref{#1}}
\newcommand{\eqns}[1]{equations~\eqref{#1}}
\newcommand{\fig}[1]{Figure~\ref{#1}}

\newcommand{\be}{\begin{equation}}
\newcommand{\ee}{\end{equation}}
\newcommand{\bear}{\begin{eqnarray}}
\newcommand{\ear}{\end{eqnarray}}
\newcommand{\nline}{\nonumber \\}
\newcommand{\f}{\frac}

\title[21~cm fluctuations using clustering wedges]{Measuring the reionization 21~cm fluctuations using clustering wedges}

\author[Raut et al.]{Dinesh Raut$^{1}$\thanks{Email:dinesh@ncra.tifr.res.in}, Tirthankar Roy Choudhury$^{1}$ and Raghunath Ghara$^{2}$\\
$^{1}$National Centre for Radio Astrophysics, TIFR, Post Bag 3, Ganeshkhind, Pune 411007, India\\
$^{2}$Department of Astronomy \& Oskar Klein Centre, AlbaNova, Stockholm University, SE-106 91 Stockholm, Sweden
}

\pubyear{2017}

\begin{document}
\label{firstpage}
\pagerange{\pageref{firstpage}--\pageref{lastpage}}
\maketitle

\begin{abstract}
One of the main challenges in probing the reionization epoch using the redshifted 21~cm line is that the magnitude of the signal is several orders smaller than the astrophysical foregrounds. One of the methods to deal with the problem is to avoid a wedge-shaped region in the Fourier $k_{\perp} - k_{\parallel}$ space which contains the signal from the spectrally smooth foregrounds. However, measuring the spherically averaged power spectrum using only modes outside this wedge (i.e., in the reionization window), leads to a bias. We provide a prescription, based on expanding the power spectrum in terms of the \emph{shifted} Legendre polynomials, which can be used to compute the angular moments of the power spectrum in the reionization window. The prescription requires computation of the monopole, quadrupole and hexadecapole moments of the power spectrum using the theoretical model under consideration and also the knowledge of the effective extent of the foreground wedge in the $k_{\perp} - k_{\parallel}$ plane. One can then calculate the theoretical power spectrum in the window which can be directly compared with observations. The analysis should have implications for avoiding any bias in the parameter constraints using 21~cm power spectrum data.
\end{abstract}


\section{Introduction}

Among the various probes of the epoch of reionization, perhaps the most promising is the study of the redshifted 21~cm signal arising from the cosmic neutral hydrogen (HI) in the intergalactic medium \citep[for reviews, see][]{2006ARA&A..44..415F,2006PhR...433..181F,2009CSci...97..841C,2012RPPh...75h6901P}. One of the main challenges in detecting the cosmological signal from the epoch of reionization using low-frequency radio telescopes is to separate out the astrophysical foregrounds \citep[see, e.g.,][]{2006PhR...433..181F,2010MNRAS.409.1647J}. These foreground signals are mainly contributed by synchrotron radiation from the Milky Way and continuum radiation of the extragalactic radio sources. In general, the magnitude of the foreground signal can surpass that of the underlying cosmological 21~cm signal by 4--5 orders \citep{2002ApJ...564..576D,2003MNRAS.346..871O,2004MNRAS.355.1053D,2008MNRAS.385.2166A}.

Although the cosmological 21~cm signal is weak, it decorrelates rapidly along the frequency direction. The foregrounds, on the other hand, are expected to be smooth functions of frequency. This distinctive feature can possibly be used to distinguish the signal from the foregrounds and hence one can hope to detect the weak cosmological signal in the observations. Various methods that have been proposed to deal with the foregrounds can be broadly separated into two categories, one in which the foreground is removed by careful modelling \citep{2005ApJ...625..575S,2006ApJ...638...20B,2006ApJ...650..529W,2008MNRAS.391..383G,2009MNRAS.398..401L,2009MNRAS.394.1575L,2009MNRAS.397.1138H,2010MNRAS.405.2492H,2011PhRvD..83j3006L,2011MNRAS.413.2103P,2012MNRAS.423.2518C,2015MNRAS.447.1973B,2015MNRAS.452.1587G}, and the second where one avoids a substantial region of the $k$-space and concentrates on a particular set of Fourier modes where the foreground is expected to be less severe \citep{2010ApJ...724..526D,2012ApJ...745..176V,2012ApJ...752..137M,2012ApJ...757..101T,2012ApJ...756..165P,2013ApJ...768L..36P,2013ApJ...770..156H,2014PhRvD..90b3018L,2014PhRvD..90b3019L,2015ApJ...804...14T}. A comparison of these two methods show that the foreground avoidance method enables better measurement of the cosmological signal at relatively large scales, while smaller scales can only be probed by foreground removal \citep{2016MNRAS.458.2928C}.

The avoidance method is based on the idea that the foregrounds are limited to a cylindrical ``wedge''-shaped region in the $k_{\perp} - k_{\parallel}$ space, where $k_{\parallel}$ ($k_{\perp}$) is magnitude of the component of the Fourier mode in the direction parallel (perpendicular) to the line of sight. In this case, the 21~cm signal from reionization can be extracted from a relatively cleaner region of the $k$-space, usually called the ``reionization window''. This technique has already been applied to analyse the 21~cm data in experiments like the PAPER\footnote{http://eor.berkeley.edu/} \citep{2015ApJ...809...61A} and MWA\footnote{http://www.mwatelescope.org/} \citep{2016ApJ...833..213P}. The same method can also be used in the case of the post-reionization 21~cm signal, e.g., for BAO surveys \citep{2016MNRAS.456.3142S}. The main technical challenge in using the avoidance method lies in ensuring that no significant foreground signal leaks into the window, e.g., from the presence of spectral structures \citep{2013ApJ...768L..36P}. There exist various sophisticated and advanced mathematical techniques for maximizing the detection of the cosmological signal in the reionization window, \citep[see, e.g.,][]{2014PhRvD..90b3018L,2014PhRvD..90b3019L}.

Even if the observational systematics and technical challenges are properly accounted for and the cosmological signal in the window is recovered with some reasonable accuracy, there exists some concern in interpreting the observations as one does not have access to the full $k$-space. This can lead to a bias in the measurement of the spherically averaged power spectrum which requires integration over \emph{all} angles. Since the peculiar velocities make the power spectrum anisotropic, calculating the spherically averaged power spectrum over a restricted region in the Fourier space can lead to a bias \citep{2014PhRvD..90b3018L,2016MNRAS.456...66J}. This may have important implications while constraining parameters related to reionization and the first stars.

Interestingly, the measurement of power spectrum (or equivalently the two-point correlation function) in a wedge-shaped region of the Fourier space has been discussed in the context of galaxy surveys \citep{2013MNRAS.435...64K}. The main aim of such studies, named as ``clustering wedges'', is to break the degeneracy between different parameters by constructing angle-averaged quantities in a restricted region. Interestingly, the presence of spectrally smooth foregrounds in the 21~cm experiments naturally produces a wedge-shaped region, hence some of the techniques developed for the clustering wedges can be extended to interpreting the data obtained through foreground avoidance studies.

The main aim of this work is to develop a method to enable unbiased comparison between theoretical model predictions and observations \emph{in presence of the foreground wedge}. In particular, we explore the possibility of using the clustering wedges to measure the power spectrum of the 21~cm signal from reionization so that the wedge bias is accounted for. The implicit assumption in this work is that the observational systematics and foregrounds are completely absent outside the wedge, and the observations have been integrated over a sufficiently long time so as to keep the noise level well below the cosmological 21~cm signal. We use previously developed semi-numerical simulations of reionization \citet{2015MNRAS.447.1806G} to model the 21~cm signal, and study the effect of the foreground wedge on the resulting power spectrum. We outline a method, based on calculating the multipoles of the 21~cm power spectrum, which can be applied to the theoretical model and allows for a fair comparison with the data.

The plan of the paper is as follows: The simulations used in the paper along with the effect of foreground wedge are discussed in Section 2. In Section 3, we discuss the clustering wedges and develop the method for comparing with the observations. The method is validated in Section 4, and we summarize our conclusions in Section 5. The cosmological parameters used are $\Omega_m = 0.32$, $\Omega_{\Lambda} = 1 - \Omega_m$, $h = 0.67$, $\Omega_b = 0.049$, $\sigma_8 = 0.83$ and $n_s = 0.96$ \citep{2014A&A...571A..16P}

\section{Simulations of the 21~cm signal and the wedge bias}

In this section, we summarize the method used for generating the reionization 21~cm signal and also discuss the concept of wedge bias following the work of \citet{2016MNRAS.456...66J}. These discussions have been included mainly for completeness and to ensure that our calculations agree with the previous work even though the methods used for simulating the 21~cm signal are different in the two studies.

\subsection{Simulations of the 21~cm signal}
\label{sec:simulations}

The simulations of the 21~cm brightness temperature used in this work is based on the one-dimensional radiative transfer simulations developed in \citet{2015MNRAS.447.1806G}. The main features of the simulation are as follows:

\begin{itemize}

\item We first generate the dark matter density and velocity fields using the publicly available code {\sc cubep}$^3${\sc m}\footnote{\tt http://wiki.cita.utoronto.ca/mediawiki/index.php/CubePM} \citep{2013MNRAS.436..540H}. We have used a cubical box of length $L_{\rm box} = 200 ~h^{-1}$ cMpc having $1728^3$ particles with the number of grid points being $3456^3$. The cosmological parameters used for the $N$-body simulation are same as listed before. This leads to a particle mass of $2 \times 10^8 ~\Msun$.

\item The dark matter haloes were identified using a runtime halofinder algorithm based on the spherical overdensity method, which leads to a minimum resolved halo mass of $\sim 2 \times 10^9 ~\Msun$. We use a subgrid model \citep{2004ApJ...609..474B} to populate the simulation box with haloes having mass below this resolution limit and above $10^8 ~\Msun$.

\item The haloes are assumed to host luminous sources that include stars and a X-ray component following a power-law spectral energy distribution (similar to what is expected from mini-QSOs). The parameters and the ionization maps used in this study are identical to what has been discussed in $S3$ model of \citet{2015MNRAS.453.3143G}. The reionization history thus obtained is consistent with the CMBR constraints as given in \citet{2016A&A...594A..13P}.

\item The baryons are assumed to follow the dark matter density field. The ionization and thermal histories of the baryons in the box are computed using a one-dimensional radiative transfer method based on earlier works of \citet{2009MNRAS.393...32T,2011MNRAS.410.1377T}. These simulations allow us to compute the spin temperature fluctuations in very early stages of reionization, along with the growth of ionized regions.

\item We also include the effect of peculiar velocities using the MM-RRM scheme of \citet{2012MNRAS.422..926M} which are essential for this work.

\end{itemize}

The \emph{redshift space} power spectrum $P(\mathbf{k})$ of the brightness temperature fluctuations $\delta T_b$ is calculated as 
\be
\avg{\delta \hat{T}_b(\mathbf{k}) ~\delta \hat{T}^*_b(\mathbf{k'})} = (2 \pi)^3~\delta_D(\mathbf{k} - \mathbf{k'})~P(\mathbf{k}),
\ee
where $\delta \hat{T}_b(\mathbf{k})$ is the Fourier transform of the fluctuations. In this work, we will be mostly working with the dimensionless power spectrum
\be
\Delta^2(\mathbf{k}) = \f{k^3 P(\mathbf{k})}{2 \pi^2}.
\ee
The spherically averaged power spectrum is obtained by averaging $\Delta^2(\mathbf{k})$ over all possible angles
\be
\Delta^2_0(k) = \f{1}{2} \int_{-1}^1 \der \mu~\Delta^2(\mathbf{k}) = \int_0^1 \der \mu~\Delta^2(\mathbf{k}),
\ee
where $\mu \equiv k_{\parallel} / k$ is the cosine of the angle between the wave vector $\mathbf{k}$ and the line of sight, and the second expression follows from the symmetry of the quantities under $\mu \to -\mu$.

\begin{figure}
    \includegraphics[width=0.45\textwidth]{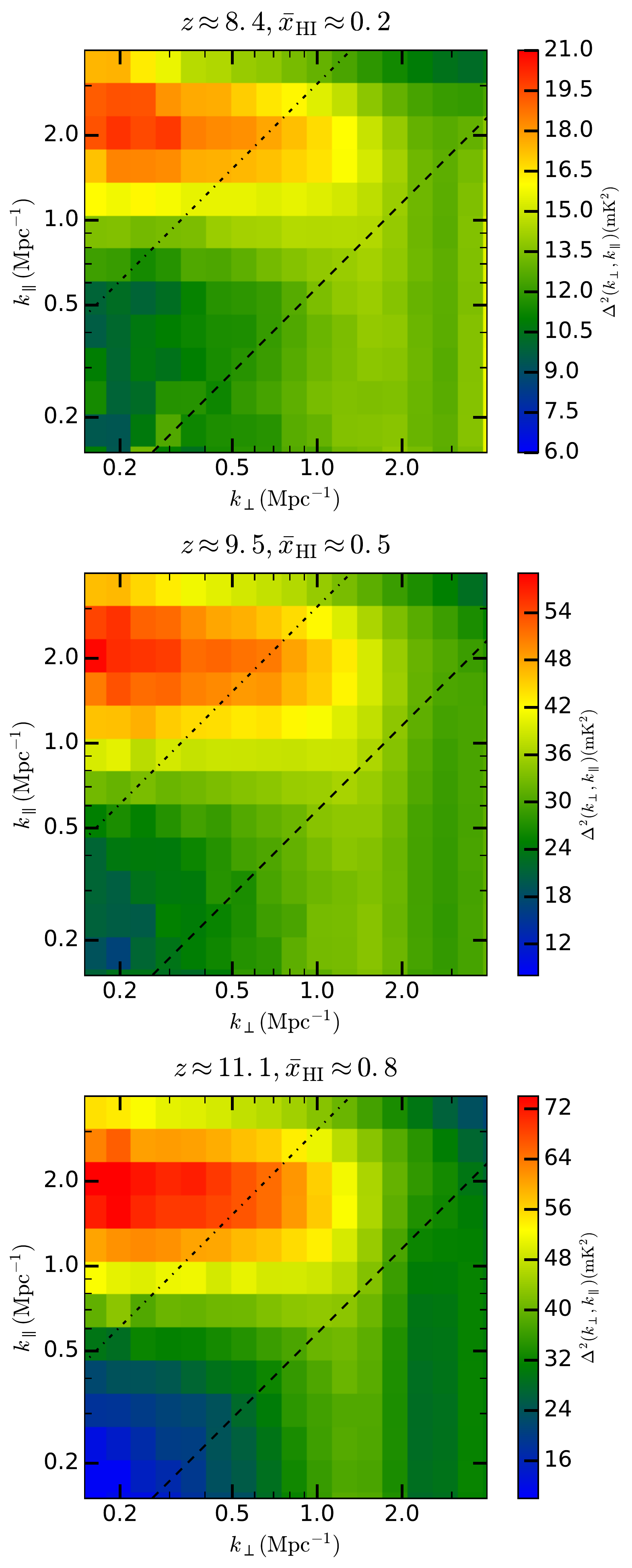}
    \caption{The power spectrum of the 21~cm brightness temperature in the $k_{\perp} - k_{\parallel}$ plane for three different redshifts. The dashed straight line corresponds to $\mu_{\rm min}=0.5$ and dot-dashed straight line corresponds to $\mu_{\rm min} = 0.95$. The regions above these lines form the reionization window, while those below the lines correspond to the foreground wedge.}
              \label{fig:cylipower}
\end{figure}

We show the power spectrum $\Delta^2(\mathbf{k})$ in the $k_{\perp} - k_{\parallel}$ plane obtained from our simulation in \fig{fig:cylipower}. The results are shown for three different redshifts. The power spectrum is highly anisotropic at relatively small scales $k \gtrsim 1$ Mpc$^{-1}$ for all the three redshifts, which arise mainly from non-linearities in the fields. At high redshifts (bottom panel), the ionization fronts have not propagated prominently into the IGM, thus the power spectrum amplitude is small at large scales $k \lesssim 0.5$ Mpc$^{-1}$. The anisotropies too are reasonably small at these scales. The large-scale amplitude of the power spectrum rises to a maximum at intermediate redshifts when $x_{\rm HI} \sim 0.5$ and decreases thereafter. This is a direct consequence of patchiness in the ionization field arising from bubbles. The large-scale anisotropies too are quite significant at these relatively low redshifts.

\subsection{The wedge bias}

\begin{figure}
    \includegraphics[width=0.5\textwidth]{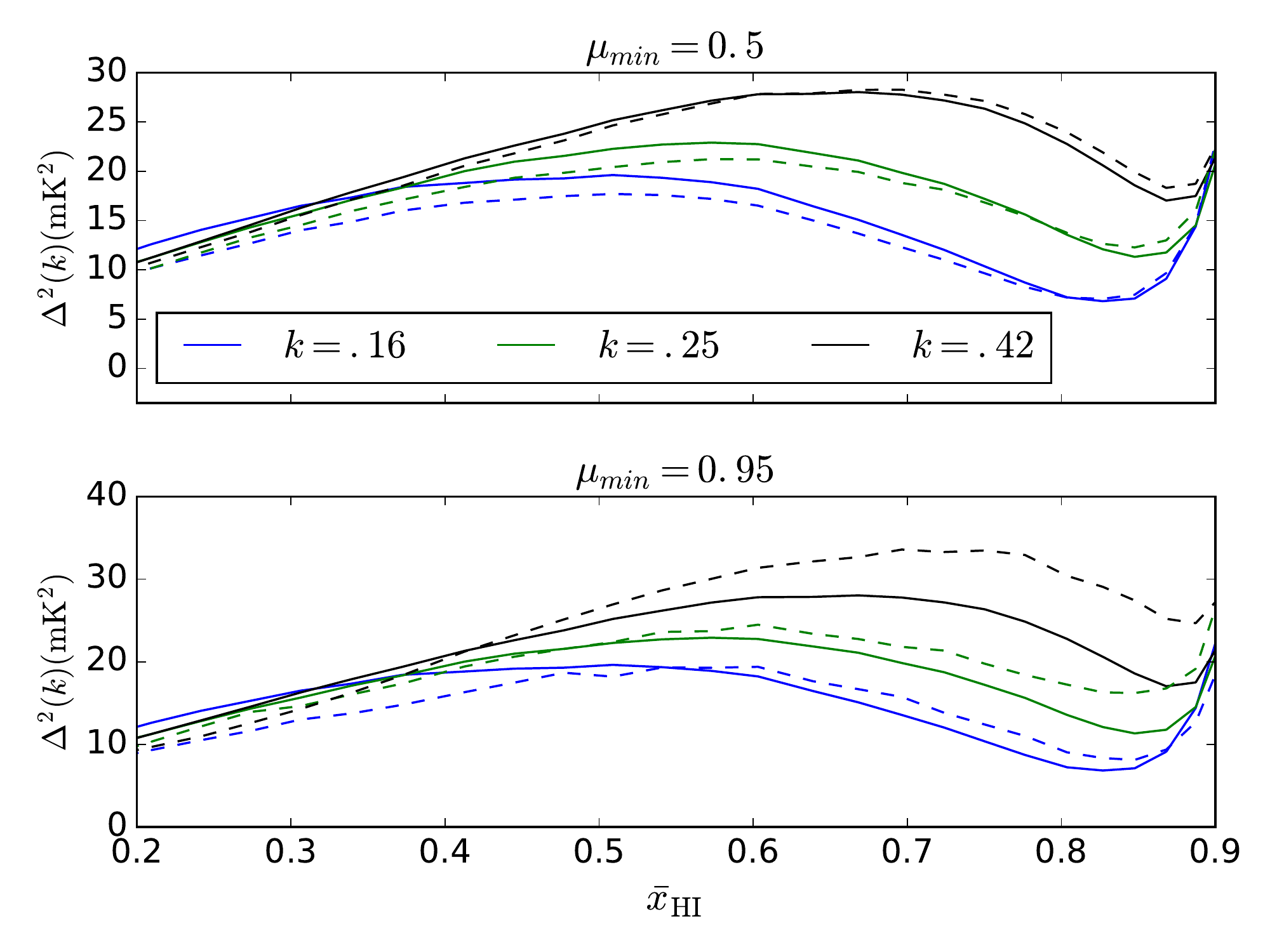}
    \caption{The difference between the spherically averaged power spectrum of the 21~cm brightness temperature obtained from the full $k$-space (solid curves) and that from the reionization window (dashed curves) as a function of the mass averaged neutral fraction $\bar{x}_{\rm HI}$ for three values of $k$ (in Mpc$^{-1}$) as indicated in the top panel. The top panel represents the case $\mu_{\rm min} = 0.5$ while the bottom panel is for $\mu_{\rm min}=0.95$.}
              \label{fig:jensen_auto}
  \end{figure}

The smoothness of the foregrounds in the frequency space ensures that they will be concentrated in the low-$k_{\parallel}$ modes. In fact, for a point source with flat spectrum at the phase centre of the telescope, the resulting power spectrum will be a delta function centred at $k_{\parallel} = 0$. For sources away from the phase centre, the foreground signal can be shown to be concentrated along a straight line in the $k_{\perp} - k_{\parallel}$ plane. Assuming a distribution of such sources in the sky (which can include diffuse radiation as well, as long as they have smooth spectra), one can show that the foreground signal is limited to a region given by \citep{2010ApJ...724..526D,2014PhRvD..89b3002D}
\be
k_{\parallel} \leq C~k_{\perp},~~
C = \sin \theta_{\rm FoV}~\f{x(z) H(z)}{c (1+z)},
\label{eq:C}
\ee
where $\theta_{\rm FoV}$ is the angular radius of the field of view (possibly set by the primary beam of individual antenna element), $x(z)$ is the comoving distance to redshift $z$ and $H(z)$ is the Hubble parameter. Note that the redshift $z$ is related to the frequency $\nu_{\rm obs}$ of observations by $1 + z = 1420~{\rm MHz} / \nu_{\rm obs}$.

The above relation allows us to define a threshold value $\mu_{\rm min}$ such that all modes in the range $-1 \leq \mu \leq -\mu_{\rm min}$ and $\mu_{\rm min} \leq \mu \leq 1$ are expected to be free from foreground contamination, which essentially defines the reionization window. This threshold value is given by 
\be
\mu_{\rm min} = \f{C}{\sqrt{1 + C^2}}.
\label{eq:mumin}
\ee
Following \citet{2016MNRAS.456...66J}, we shall present results for two values $\mu_{\rm min} = 0.5$ and $0.95$, the latter corresponding to a terribly pessimistic case where one has access to only a few modes for computing the HI power spectrum. For reference, the $\mu_{\rm min} = 0.5$ would correspond to a $\theta_{\rm FoV} \sim 10^{\circ}$ at $z \sim 8$, while $\mu_{\rm min} = 0.95$ would correspond to a $\theta_{\rm FoV} \gtrsim 60^{\circ}$ at the same redshift \citep{2016MNRAS.456...66J}. We should mention here that the effective value of $\mu_{\rm min}$ can be larger than the theoretically expected value given by \eqns{eq:C} and \eqref{eq:mumin} when the foregrounds spill into the window (e.g., because of structures in the frequency). On the other hand, the effective $\mu_{\rm min}$ can be smaller than the theoretical expectation if the reionization window can be enlarged via advances statistical methods \citep{2014PhRvD..90b3019L}. For the purpose of this paper, we assume that all the Fourier modes in the window are foreground-free.

Given $\mu_{\rm min}$, one can calculate the spherically averaged 21~cm power spectrum \emph{in the window} as
\be
\Delta^2_{0, {\rm win}}(k) = \f{1}{1 - \mu_{\rm min}} \int_{\mu_{\rm min}}^1 \der \mu~\Delta^2(\mathbf{k}).
\ee
When the power spectrum is isotropic $\Delta^2(\mathbf{k}) = \Delta^2(k)$, the above relation reduces to $\Delta^2_{0, {\rm win}}(k) = \Delta^2_0(k)$ as expected. However, these two quantities will not be equal when the power spectrum becomes anisotropic (e.g., in presence of peculiar velocities), which would consequently give rise to the wedge bias.

This bias for our simulation box is shown in \fig{fig:jensen_auto} where we plotted both $\Delta_0^2(k)$ and $\Delta^2_{0, {\rm win}}(k)$ as a function of the mass averaged neutral fraction $\bar{x}_{\rm HI}$ for three different values of $k$. The top panel is for $\mu_{\rm min} = 0.5$ while the bottom is for $\mu_{\rm min} = 0.95$. Our results are in agreement with those of \citet{2016MNRAS.456...66J}, even though the simulations used in these two works are different [we remind that \citet{2016MNRAS.456...66J} have used a semi-numerical method based on \citet{2007ApJ...654...12Z,2009MNRAS.394..960C,2014MNRAS.443.2843M}, while we use the one described in \citet{2015MNRAS.447.1806G}]. We can see from the figure that the power spectra calculated in the window are different from the true power spectra, the difference being larger for relatively larger scales (i.e., smaller values of $k$). Also, as expected, the difference is larger for higher values of $\mu_{\rm min}$ as more number of modes are discarded.

\section{Clustering wedges}

The redshift space power spectrum can be decomposed in the basis of Legendre polynomials as done in case of galaxy redshift surveys \citep[see e.g.,][]{1992ApJ...385L...5H,1995MNRAS.275..515C}. The effect of the wedge bias can accounted for by using the so-called clustering wedges as was introduced also in the context of galaxy surveys by \citet{2012MNRAS.419.3223K}. One can begin by expanding the anisotropic power spectrum in the redshift space in terms of the Legendre polynomials ${\cal P}_l(\mu)$ as
\be
\Delta^2(\mathbf{k}) \equiv \Delta^2(k, \mu) = \sum_{l~{\rm even}} \Delta^2_l(k)~{\cal P}_l(\mu),
\ee
where the symmetry under $\mu \to -\mu$ ensures that only even $l$'s contribute to the sum. In the rest of the paper, we shall be concerned with only the even multipoles, though most of the discussion can be easily generalized to cases where the symmetry is not present. 
Such a splitting has been used in case of models of reionization \citep[see e.g.,][]{2013MNRAS.434.1978M,2016MNRAS.456.2080M}. The multipoles $\Delta^2_l(k)$ of the power spectrum can be written as
\be
\Delta^2_l(k) = (2 l +1)\int_{0}^1 \der \mu~\Delta^2(k, \mu)~{\cal P}_l(\mu), ~~l = 0,2,4,\ldots
\label{eq:Delta_sq_l}
\ee
Note that the $l=0$ term corresponds to the spherically averaged power spectrum defined earlier. 

It is straightforward to show that the spherically averaged power spectrum $\Delta^2_{0, {\rm win}}(k)$, or the clustering wedge, evaluated in the window $\mu_{\rm min} \leq \mu \leq 1$ is given in terms of the $\Delta^2_l(k)$ as
\be
\Delta^2_{0, {\rm win}}(k) = \sum_{l~{\rm even}} \Delta^2_l(k) \left[\f{1}{1 - \mu_{\rm min}} \int_{\mu_{\rm min}}^1 \der \mu~{\cal P}_l(\mu)\right].
\ee
In the linear (or quasi-linear) models of redshift space distortions, the only values of $l$ that contribute to the sum are $0, 2, 4$ \citep{2012MNRAS.422..926M}. In that case, one can write an explicit expression for the clustering wedge as
\bear
\Delta^2_{0, {\rm win}}(k) &=& \Delta^2_0(k) + \f{1}{2} \mu_{\rm min} \left(1 + \mu_{\rm min}\right) \Delta^2_2(k) 
\nline
&+& \f{1}{8} \mu_{\rm min} \left(1 + \mu_{\rm min}\right) \left(7 \mu_{\rm min}^2 - 3\right) \Delta^2_4(k).
\label{eq:deltasq_0_win}
\ear
The above expression shows that the wedge bias is contributed by the higher order multipoles. In other words, the incomplete $k$-space coverage leads to mixing of different multipoles which must be accounted for while comparing theoretical models with the data. Note that the bias vanishes for $\mu_{\rm min} \to 0$ which corresponds to the case where one has access to all the $k$-modes for calculating the reionization power spectrum.

\begin{figure*}
    \includegraphics[width=1.0\textwidth]{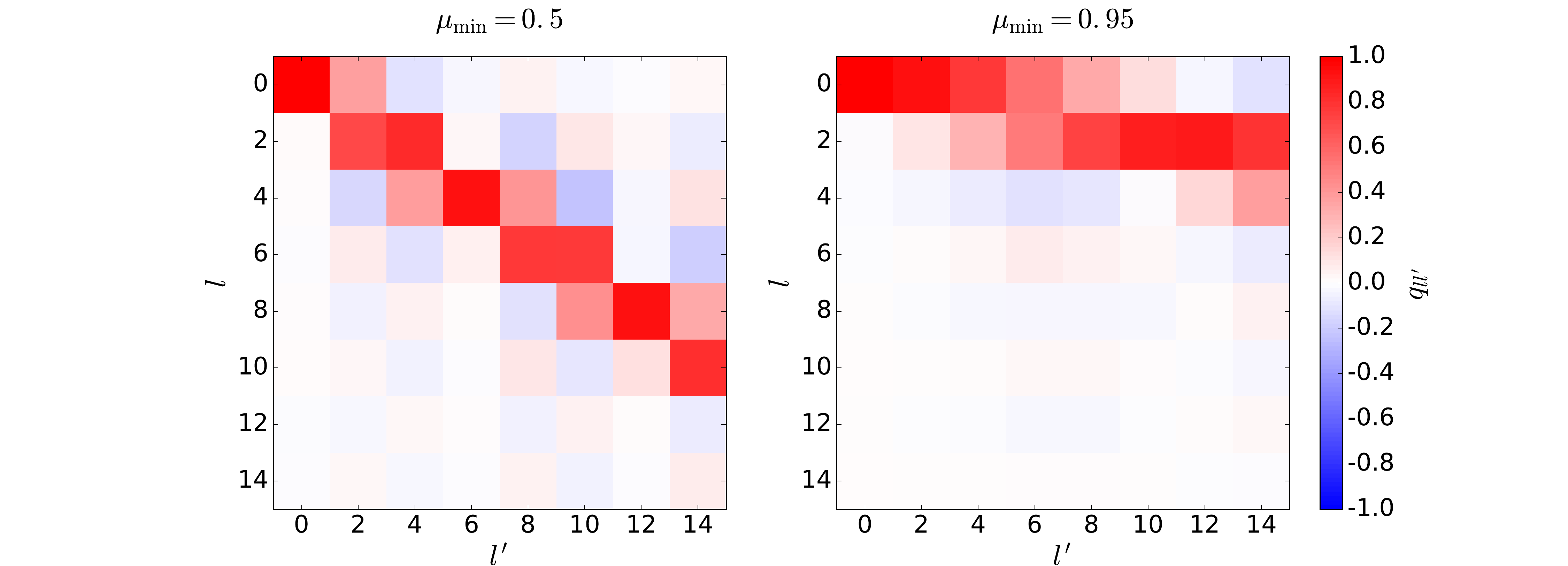}
    \caption{The bias matrix $q_{l l'}$ defined in \eqn{eq:matrix_q} for two values of $\mu_{\rm min} = 0.5$ (left panel) and $0.95$ (right panel).} 
              \label{fig:plot_matrix_q}
  \end{figure*}

In order to get some further insight into the origin of the wedge bias, let us write the brightness temperature fluctuations in Fourier space as \citep{2005ApJ...624L..65B}
\be
\delta \hat{T}_b(\mathbf{k}) = \bar{T}_b \left[\delta_{\rm HI}(\mathbf{k}) + \mu^2 \delta_b(\mathbf{k})\right],
\label{eq:delta_Tb_k}
\ee
where $\bar{T}_b$ is the mean brightness temperature, and $\delta_b$ ($\delta_{\rm HI}$) is the density contrast in baryons (HI). We have assumed the HI spin temperature $T_S$ to much larger than the radiation temperature and hence ignored the fluctuations in $T_S$. This assumption is reasonable for $\bar{x}_{\rm HI} \lesssim 0.9$ \citep{2015MNRAS.447.1806G}, and also simplifies the subsequent discussion. However, it is straightforward to extend the discussions to cases where the fluctuations in $T_S$ cannot be ignored.

Under the quasi-linear approximation, the multipole moments of the brightness temperature power spectrum are given by
\bear
\Delta^2_0(k) &=& \bar{T}_b^2 \left[\f{1}{5} \Delta^{2 (r)}_{bb}(k) + \f{2}{3} \Delta^{2 (r)}_{b, {\rm HI}}(k) + \Delta^{2 (r)}_{\rm HI, HI}(k) \right],
\nline
\Delta^2_2(k) &=& \bar{T}_b^2 \left[\f{4}{7} \Delta^{2 (r)}_{bb}(k) + \f{4}{3} \Delta^{2 (r)}_{b, {\rm HI}}(k) \right],
\nline
\Delta^2_4(k) &=& \bar{T}_b^2 \f{8}{35} \Delta^{2 (r)}_{bb}(k),
\ear
where $\Delta^{2 (r)}_{bb}(k)$ and $\Delta^{2 (r)}_{\rm HI, HI}(k)$ are the \emph{real space} power spectra of baryonic and HI fluctuations respectively and $\Delta^{2 (r)}_{b, {\rm HI}}(k)$ is the corresponding cross power spectrum. Note that the cross term $\Delta^{2 (r)}_{b, {\rm HI}}(k)$ can take negative values while $\Delta^{2 (r)}_{bb}(k)$ and $\Delta^{2 (r)}_{\rm HI, HI}(k)$ are always positive. Given the above relations, we can write the wedge bias $b_{\rm wedge}(k)$ as
\bear
b^2_{\rm wedge}(k) &\equiv& \Delta^2_{0, {\rm win}}(k) - \Delta^2_0(k)
\nline
&=& \bar{T}_b^2 \mu_{\rm min} \left(1 + \mu_{\rm min}\right) \left[\f{1 + \mu_{\rm min}^2}{5} \Delta^{2 (r)}_{bb}(k) + \f{2}{3} \Delta^{2 (r)}_{b, {\rm HI}}(k) \right].
\nline
\ear
The first term in square brackets in the above expression is always positive, while the second term can be either positive or negative depending on the nature of the cross correlation. If we restrict to relatively large scales, then the HI fluctuations follow the baryonic fluctuations at very early stages of reionization and hence $\Delta^{2 (r)}_{b, {\rm HI}}(k)$ is positive. In that case we expect the bias to be positive, as is seen in \fig{fig:jensen_auto}. On the other hand, at later stages of reionization, the inside-out nature of the process makes the correlation negative at large scales, and hence the bias becomes negative. 

Note that the power spectrum estimated from the simulation box can contain multipoles of higher $l > 4$ order  than what is predicted by the quasi-linear model, even if we concentrate only on large scales. This is because the HI fluctuations $\delta_{\rm HI}$ in \eqn{eq:delta_Tb_k} are not necessarily linear and can be $\gtrsim 1$ when $\bar{x}_{\rm HI} \sim 0.5$. In addition, we find significant higher order multipoles in our simulation box arising from numerical effects due to finite box size\footnote{The higher order multipoles could also arise from the Alcock-Paczynski effect \citep{1979Natur.281..358A}, however, this is not relevant for the present study as we assume that the values of the cosmological parameters are known a priori.}. We minimize the effect of box size by considering only modes $k \gtrsim 10 \pi / L_{\rm box} \sim 0.1$ Mpc$^{-1}$. However, the non-linearities in the ionization field can be significant even at large scales and in that case the relations obtained using the quasi-linear approximation are not valid in the strict sense.

One can extend the definition of the clustering wedge $\Delta^2_{0, {\rm win}}(k)$ to higher multipoles. In order to do this, first note that the Legendre polynomials ${\cal P}_l(\mu)$ do \emph{not} form an orthogonal basis in the interval $\mu_{\rm min} \leq \mu \leq 1$. A more convenient basis to work with is the one formed by the shifted Legendre polynomials \citep[see, e.g.,][]{1970hmfw.book.....A}, which in our case turns out to be
\be
\tilde{\cal{P}}_l(\mu) = {\cal P}_l\left(\f{\mu - \mu_{\rm min}}{1 - \mu_{\rm min}}\right), ~~\mu_{\rm min} \leq \mu \leq 1, ~~ l = 0,2,4,\ldots
\ee
The above relation essentially corresponds to a shift in the interval $[0, 1] \to [\mu_{\rm min}, 1]$ through an appropriate scaling. It is straightforward to show that the new polynomials satisfy the orthogonality condition
\be
\int_{\mu_{\rm min}}^1 \der \mu~\tilde{\cal{P}}_l(\mu)~\tilde{\cal{P}}_{l'}(\mu) = \f{1 - \mu_{\rm min}}{2 l + 1} ~ \delta_{l l'}.
\ee

One can now expand the redshift space power spectrum defined in the reionization window in terms of the shifted Legendre polynomials as
\be
\Delta^2(k, \mu) = \sum_{l~{\rm even}} \Delta^2_{l, {\rm win}}(k)~\tilde{{\cal P}}_l(\mu), ~~ \mu_{\rm min} \leq \mu \leq 1.
\ee
The above relation can be inverted to obtain the multipoles as
\be
\Delta^2_{l, {\rm win}}(k) = \f{2 l +1}{1 - \mu_{\rm min}} \int_{\mu_{\rm min}}^1 \der \mu~\Delta^2(k, \mu)~\tilde{{\cal P}}_l(\mu), ~~l = 0,2,4,\ldots
\label{eq:Delta_sq_l_win}
\ee
Combining the above relation with \eqn{eq:Delta_sq_l}, we can show that the multipoles in the window are related to the true multipoles as
\be
\Delta^2_{l, {\rm win}}(k) = \sum_{l'~{\rm even}} q_{ll'}~\Delta^2_{l'}(k),
\label{eq:Delta_sq_l_win_Delta_sq_l}
\ee
where $q_{l l'}$ is the \emph{bias matrix} and is given by
\bear
q_{ll'} &=& \f{2 l + 1}{1 - \mu_{\rm min}} \int_{\mu_{\rm min}}^1 \der \mu~\tilde{{\cal P}}_l(\mu)~{\cal P}_{l'}(\mu)
\nline
&=& \f{2 l + 1}{1 - \mu_{\rm min}} \int_{\mu_{\rm min}}^1 \der \mu~{\cal P}_l\left( \f{\mu - \mu_{\rm min}}{1 - \mu_{\rm min}} \right)~{\cal P}_{l'}(\mu).
\label{eq:matrix_q}
\ear
When $\mu_{\rm min} = 0$, the bias matrix reduces to the the unit matrix $q_{l l'} = \delta_{l l'}$, while for other values of $\mu_{\rm min}$ it quantifies the bias present in the quantities computed using modes only within the reionization window.

Given the value of $\mu_{\rm min}$ appropriate for the experiment, the matrix $q_{l l'}$ needs to be evaluated only once. The plot of the matrix for two values of $\mu_{\rm min}$ is shown in \fig{fig:plot_matrix_q}. The first point to note is that when $l' = 0$, we have $q_{l 0} = \delta_{l 0}$, which is also obvious from the definition of the bias matrix in \eqn{eq:matrix_q}. This immediately implies that the higher order multipoles $\Delta^2_{l, {\rm win}}(k),~l \geq 2$ in the window do not contain any contribution from the true monopole $\Delta^2_0(k)$ and are only dependent on the higher order multipoles. Thus the detection of $\Delta^2_{l, {\rm win}}(k),~ l\geq 2$ in the reionization window would imply presence of line of sight anisotropies in the power spectrum.

We can also see from the figure that the relative amplitudes of the off-diagonal terms increase for the higher value of $\mu_{\rm min}$, which would effectively result in a higher wedge bias. Another interesting point to note is that even if the true power spectrum $\Delta^2(k)$ does not contain multipoles higher than the quadrupole $l=4$ (as would be the case in absence of non-linearities and box size effects), the power spectrum in the window can still have higher order multipoles because of the off-diagonal terms in $q_{l l'}$. We shall return to this point later in the paper.

\begin{figure*}
    \includegraphics[width=0.9\textwidth]{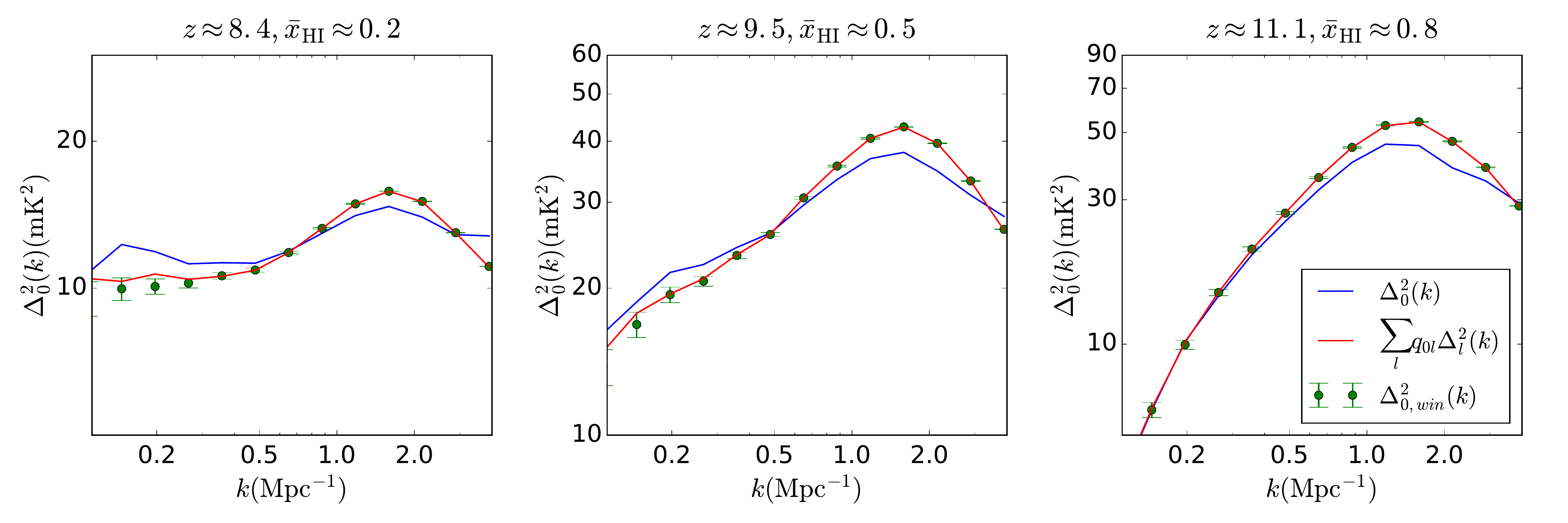}
    \caption{The spherically averaged power spectrum $\Delta^2_0(k)$ calculated over the full $k$-space (blue curves) and the power spectrum $\Delta^2_{0, {\rm win}}(k)$ calculated in the reionization window (points with error-bars) for three redshifts $z = 8.4, 9.5, 11.1$ and $\mu_{\rm min} = 0.5$. The values of the mass averaged neutral fraction $\bar{x}_{\rm HI}$ are mentioned in the respective panels. The red curves show the spherically averaged power spectrum $\sum_{l'=0,2,4} ~q_{0l'}~\Delta^2_{l'}(k)$ in the window constructed using the clustering wedge \eqn{eq:deltasq_0_win} (or equivalently using \eqn{eq:Delta_sq_l_win_Delta_sq_l} for $l = 0$ and the series terminated at $l' = 4$).} 
              \label{fig:clustering_wedge_recovery_3panels_mumin0p5}
  \end{figure*}

\begin{figure*}
    \includegraphics[width=0.9\textwidth]{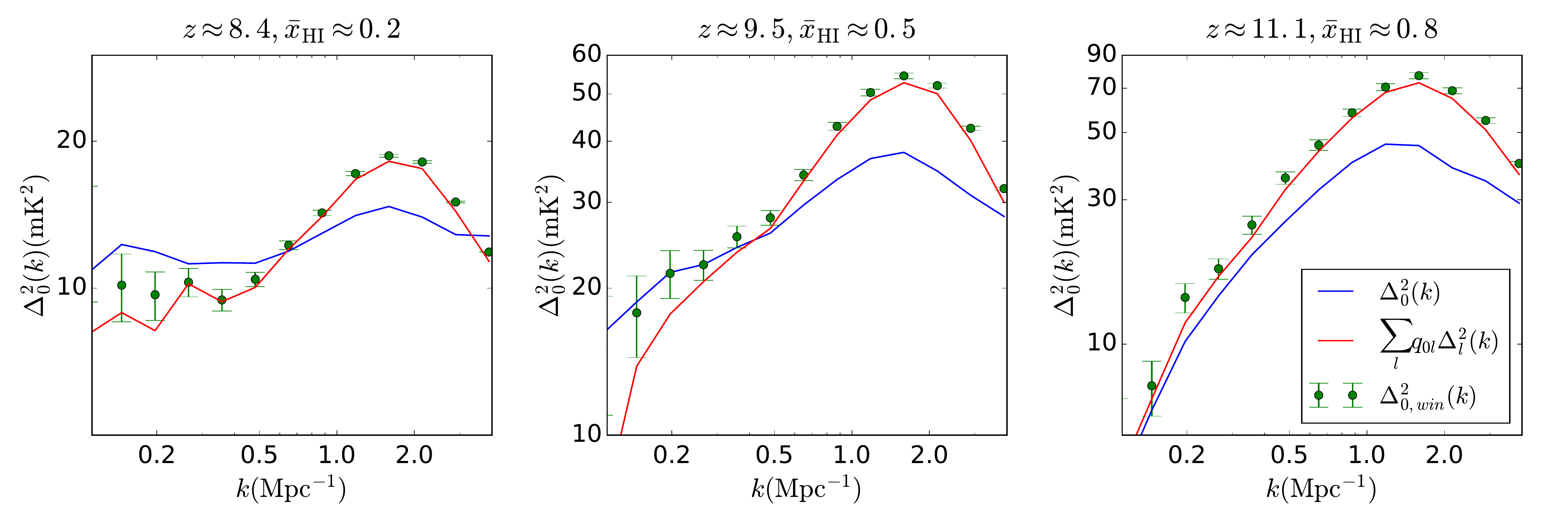}
    \caption{Same as \fig{fig:clustering_wedge_recovery_3panels_mumin0p5} but for $\mu_{\rm min} = 0.95$.} 
              \label{fig:clustering_wedge_recovery_3panels_mumin0p95}
  \end{figure*}

\section{Model comparison in presence of the wedge}

As should be obvious from the discussions above, the presence of the wedge bias implies that proper interpretation of the observations while using the foreground avoidance techniques require careful treatment of the clustering wedges. There are various possible approaches in dealing with this issue. The first obvious method would be to apply the same $\mu$-space restrictions in the simulations as one would expect in the actual observations. In this approach, the mock observations created from the simulations should automatically account for the foreground wedge and thus allow for fair comparison with the data.

However, one could envisage possible situations where the theoretical models do not allow for straightforward incorporation of the wedge effects. For example, if the simulations are limited by box size, restricting to a small range in $\mu$-space may lead to unrealistically small number of modes left to work with, which in turn would make the comparison with data very difficult. Also, it might be possible that incorporating the wedge effects in the simulation effectively slows down the calculations, which in turn will affect the parameter estimation methods that require evaluation of the power spectra for a large number of model parameters. As another example, there may not exist any obvious and uncomplicated method of incorporating the wedge effects while working with (semi-)analytical models. In such situations, one can still calculate the power spectrum multipoles $\Delta^2_{l, {\rm win}}(k)$ in the reionization window from the true multipoles $\Delta^2_l(k)$ using \eqn{eq:Delta_sq_l_win_Delta_sq_l}. This method would allow for a fair comparison with the data in presence of the wedge \emph{without} explicitly incorporating the wedge effects in the calculations. Instead, one only needs to compute the higher order multipoles $\Delta^2_l(k)$ in addition to the spherically averaged power spectrum from the theoretical model.

\begin{figure}
\includegraphics[width=0.5\textwidth]{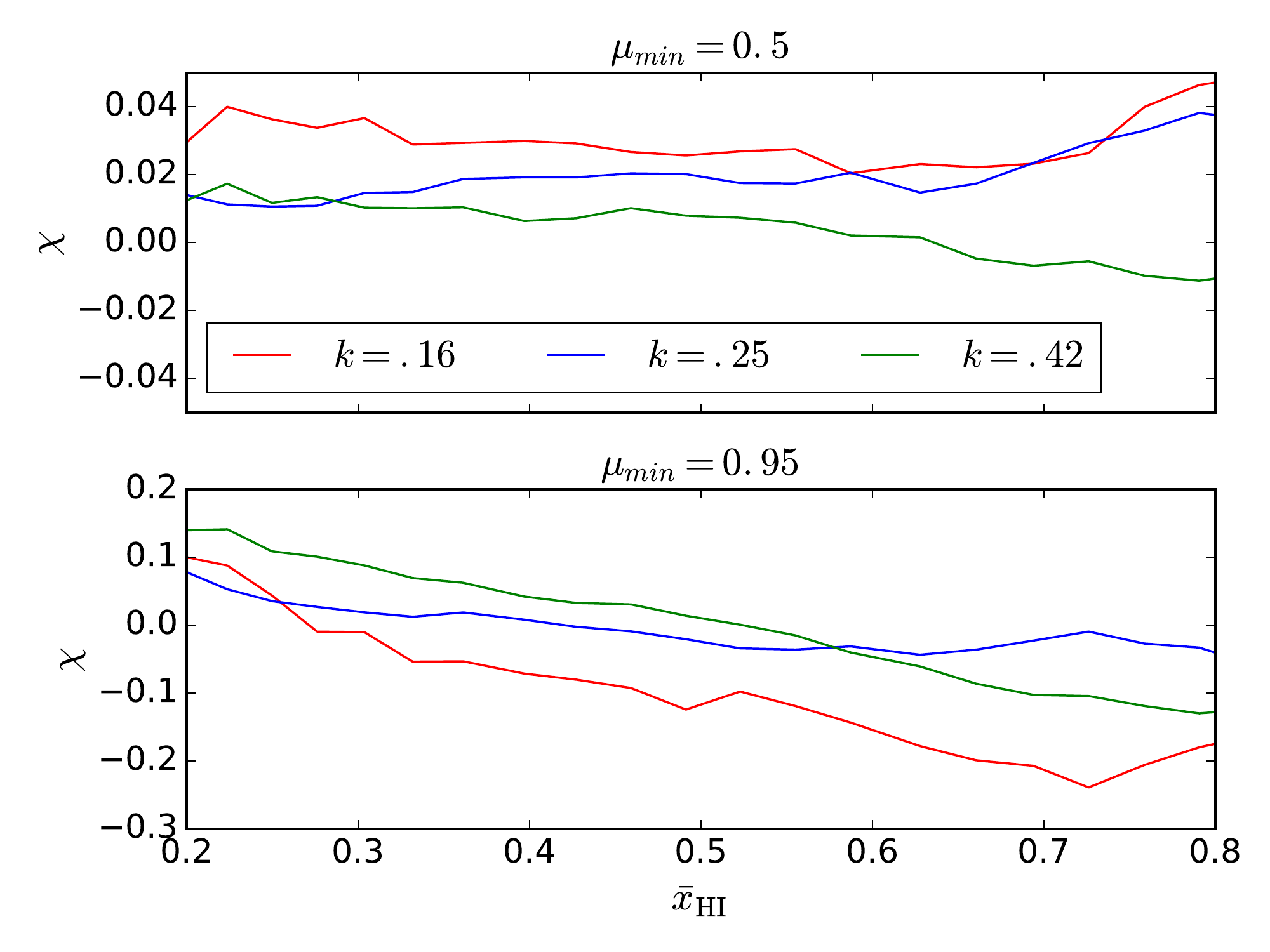}
\caption{The difference between the clustering wedge $\sum_{l'=0,2,4} ~q_{0l'}~\Delta^2_{l'}(k)$ and the spherically averaged power spectrum $\Delta^2_{0, {\rm win}}(k)$ calculated in the window, divided by the standard deviation of the dimensionless power spectrum [see \eqn{eq:chi_k}] as a function of the reionization history, i.e., $\bar{x}_{\rm HI}$. The top panel is for $\mu_{\rm min} = 0.5$ while the bottom one is for $\mu_{\rm min}=0.95$.}
\label{fig:jensen_expt432_compare_together} 
\end{figure}  

The only issue in using \eqn{eq:Delta_sq_l_win_Delta_sq_l} to estimate the $\Delta^2_{l, {\rm win}}(k)$ is to decide on how many terms to retain in the summation involving $l'$. In the quasi-linear models, the series naturally terminates at $l' = 4$, however, the presence of non-linearities could allow for higher order terms to be significant and thus making the calculations less reliable. In that case, the convergence of the series in \eqn{eq:Delta_sq_l_win_Delta_sq_l} would depend on the amplitudes of the $\Delta_{l'}^2(k)$ and also the matrix $q_{l l'}$. We can see from \fig{fig:plot_matrix_q} that the values of $q_{ll'} \lesssim 0.1$ for $l' > 4$ for $l = 0, 2$ when we take $\mu_{\rm min} = 0.5$. Hence we expect the series in \eqn{eq:Delta_sq_l_win_Delta_sq_l} to show reasonable convergence if we terminate at $l' = 4$ (unless the higher multipoles $\Delta_{l'}^2(k), l' > 4$ are significantly larger in magnitude than the lower $l' \leq 4$ ones). The situation is less favourable for the $\mu_{\rm min} = 0.95$ case where we find that, even for $l = 0$, the values of $q_{ll'} \sim 0.4$ for $l' \leq 8$. Hence the convergence of the series will depend on how quickly the higher multipoles $\Delta_{l'}^2(k)$ decrease with increasing $l'$.

In the following, we study this issue in slightly more detail using the simulations discussed in Section \ref{sec:simulations}. The simulations, in principle, incorporate non-linearities in the ionization and density fields in relevant scales and hence any conclusions drawn would not depend on the assumptions related to quasi-linear approximations.

In \fig{fig:clustering_wedge_recovery_3panels_mumin0p5} we plot the ``true'' spherically averaged power spectrum $\Delta_0^2(k)$ for three redshifts with $\mu_{\rm min} = 0.5$ (shown by blue curves), along with the corresponding quantity $\Delta_{0, {\rm win}}^2(k)$ calculated in the reionization window (shown by green points with error-bars) by accounting for only modes with $\mu \geq \mu_{\rm min}$. The error-bars represent the 1--$\sigma$ statistical error $\sigma_{\Delta^2_{0, {\rm win}}(k)}$ estimated from the variance within each $k$-bin. The fact that the blue curves do not follow the green points is a demonstration of the wedge bias discussed earlier, and hence once cannot use $\Delta_0^2(k)$ calculated from the simulations to compare with the observations in the reionization window. 

\begin{figure*}
    \includegraphics[width=0.9\textwidth]{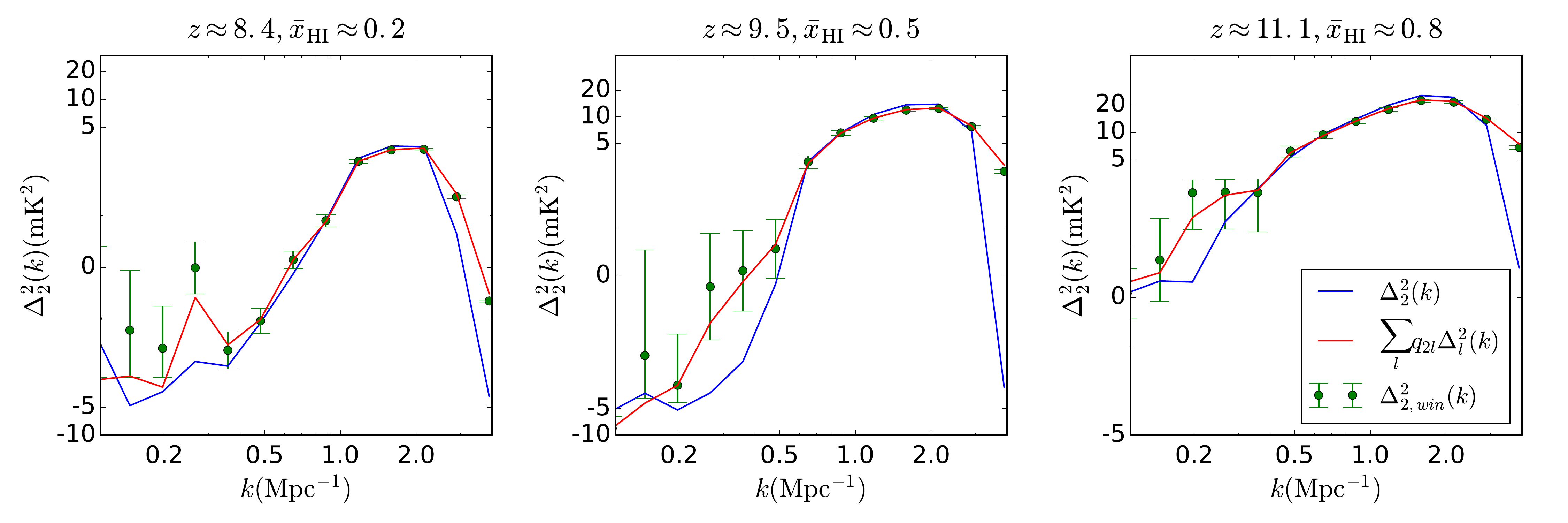}
    \caption{The quadrupole moment of the power spectrum $\Delta^2_2(k)$ calculated over the full $k$-space (blue curves) and the quadrupole moment $\Delta^2_{2, {\rm win}}(k)$ calculated in the reionization window (points with error-bars) for three redshifts $z = 8.4, 9.5, 11.1$ and $\mu_{\rm min} = 0.5$. The values of the mass averaged neutral fraction $\bar{x}_{\rm HI}$ are mentioned in the respective panels. The red curves show the spherically averaged power spectrum $\sum_{l'=0,2,4} ~q_{2l'}~\Delta^2_{l'}(k)$ in the window constructed using the clustering wedge \eqn{eq:Delta_sq_l_win_Delta_sq_l}.} 
\label{fig:clustering_wedge_recovery_higher_mumin0p5}
\end{figure*}

Now let us calculate the next two (even) multipoles $\Delta_2^2(k)$ and $\Delta_4^2(k)$ from the simulation box (using the full $k$-space) and subsequently use \eqn{eq:deltasq_0_win} to estimate the clustering wedge $\sum_{l'=0,2,4} ~q_{0l'}~\Delta^2_{l'}(k)$ (this is same as using \eqn{eq:Delta_sq_l_win_Delta_sq_l} for $l = 0$ and the series terminated at $l' = 4$). The results are shown by the red curves.  It is clear that these curves agree quite well with the points which are the true values $\Delta_{0, {\rm win}}^2(k)$ in the simulation. Any difference between the clustering wedge and the true values are within the statistical errors. Thus, using the clustering wedge of \eqn{eq:deltasq_0_win} allows for a proper comparison of the theoretical models with the data. Note that this method does \emph{not} require incorporating the wedge effects in the theoretical models and can be performed by computing the multipoles using the full $\mu$-space and the matrix $q_{l l'}$.

The results for $\mu_{\rm min} = 0.95$ are shown in \fig{fig:clustering_wedge_recovery_3panels_mumin0p95}. In this case, one has access to much less number of modes in the reionization window, hence the statistical errors are larger than in the previous case. This is clearly manifested in the larger size of the error-bars for $\Delta_{0, {\rm win}}^2(k)$ calculated from the simulation box. We also find that, in general, the spherically averaged power spectrum $\Delta_0^2(k)$ calculated using all the modes (blue curve) does not agree with $\Delta_{0, {\rm win}}^2(k)$. The red curve, which represents the clustering wedge $\sum_{l'=0,2,4} ~q_{0l'}~\Delta^2_{l'}(k)$, has a better agreement with the true value $\Delta_{0, {\rm win}}^2(k)$. The exception to this agreement can be seen at large case for $\bar{x}_{\rm HI} \approx 0.5$ (middle panel). In this case, the clustering wedge estimates deviate from the true value because of two reasons: first is that the matrix elements $q_{0 l'}$ are not negligible for $l' > 4$, and second is that the non-linearities in the $\delta_{\rm HI}$ are more significant when $\bar{x}_{\rm HI} \sim 0.5$. Hence we conclude that terminating the series in \eqn{eq:Delta_sq_l_win_Delta_sq_l} at $l' = 4$ gives less accurate results for the $\mu_{\rm min} = 0.95$ case, and one should attempt to retain higher order terms. In our simulation box, including the higher order terms leads to predictions that are somewhat noisy at large scales, hence we do not attempt to do so in this work. In general, we found that retaining terms up to $l' = 4$ works well for $\mu_{\rm min} \lesssim 0.8$.

In order to compare how the clustering wedge compares with the power spectrum in the reionization window, we define the quantity
\be
\chi(k) = \f{1}{\sigma_{\Delta^2_{0, {\rm win}}(k)}} \left[ \sum_{l=0,2,4} q_{0l} \Delta^2_l(k) - \Delta^2_{0, {\rm win}}(k)\right]
\label{eq:chi_k}
\ee
which essentially measures the difference between the window power spectrum and the clustering wedge normalized by the standard deviation. The plot of $\chi(k)$ for three values of $k$ as a function of $\bar{x}_{\rm HI}$ is shown in \fig{fig:jensen_expt432_compare_together}. The top and bottom panels show the results for $\mu_{\rm min}=0.5$ and $\mu_{\rm min}=0.95$, respectively. As seen from the plots, for the $\mu_{\rm min}=0.5$ case, there is very good agreement between the power spectrum in the window and the clustering wedge as the magnitude of $\chi$ is always less than $\sim 5\%$, showing that the differences are much less than the statistical errors. On the other hand, the agreement is relatively poor for $\mu_{\rm min}=0.95$, however the magnitude of $\chi$ is still $\lesssim 30\%$.

In addition to the monopole term, one can also compute the higher order multipoles in the wedge using higher order clustering wedges \eqn{eq:Delta_sq_l_win_Delta_sq_l}. The results for the quadrupole $l = 2$ are shown in \fig{fig:clustering_wedge_recovery_higher_mumin0p5}. The conclusions are similar to those discussed above. We find that the quadrupole moment $\Delta^2_{2, {\rm win}}(k)$ calculated in the window (green points with error-bars) has a decent agreement with that calculated using the clustering wedge \eqn{eq:Delta_sq_l_win_Delta_sq_l} retaining terms only up to $l' = 4$ in the series (red curves). In contrast, the quadrupole calculated using all the modes (blue curves) do not agree that well with the green points. We also attempted the same for $\mu_{\rm min} = 0.95$, however, the agreement between $\Delta^2_{2, {\rm win}}(k)$ and the clustering wedge is quite poor in that case because of the non-negligible values of the matrix elements $q_{2 l'},~ l' > 4$ and the non-linearities in the $\delta_{\rm HI}$ field.

Given the above results, we suggest the following procedure for comparing the theoretical models with observations in presence of the foreground wedge: 
\begin{enumerate}
\item calculate the moments $\Delta^2_l(k), ~l=0,2,4$ using the full $k$-space from the theoretical model under consideration,
\item given the observational parameters, calculate $\mu_{\rm min}$ and hence the matrix $q_{l l'}$, and
\item calculate the clustering wedges $\sum_{l' = 0,2,4} q_{l l'} \Delta^2_{l'}(k)$. 
\end{enumerate}
This final product can be compared with the multipoles $\Delta^2_{l, {\rm win}}(k)$ calculated in the reionization window. The method gives accurate results for the spherically averaged power spectrum $l = 0$ and $\mu_{\rm min} \lesssim 0.8$. Since most simulations and/or semi-analytical models predict the higher order multipoles along with the monopole, this should be a straightforward extension to any models for constraining the reionization parameters.

One can also attempt a slightly different method while interpreting the quantities calculated in the window. For example, we can ask whether it is possible to construct an unbiased estimator for the spherically averaged power spectrum using only the modes available in the reionization window. In principle, one can invert \eqn{eq:Delta_sq_l_win_Delta_sq_l} and write it as
\be
\Delta^2_{l}(k) = \sum_{l'~{\rm even}} Q_{ll'}~\Delta^2_{l', {\rm win}}(k),
\label{eq:Delta_sq_l_Delta_sq_l_win}
\ee
where $Q_{l l'}$ is simply the inverse matrix of $q_{l l'}$
\be
\sum_{n~ {\rm even}} Q_{l n}~q_{n l'} = \delta_{l l'}, ~~ l, l' = 0,2,4,\ldots.
\ee
It may seem from \eqn{eq:Delta_sq_l_Delta_sq_l_win} that if one is able to measure the multipoles $\Delta^2_{l', {\rm win}}(k)$ in the window, it should be possible to construct the true multipoles $\Delta^2_l(k)$, which then provides a direct method of comparing with theoretical models.

However, this method leads to several difficulties. Note that $\Delta^2_{l, {\rm win}}(k)$ is non-zero for all values of $l$ (even if $\Delta^2_{l}(k)$ terminates at $l=4$), hence one needs to formally sum over an infinite number of terms to obtain the unbiased estimator. In practise, one has to check if the series converges when a reasonably small number of terms included in the expression. The convergence will depend on the dependence of $\Delta^2_{l', {\rm win}}(k)$ on $l'$ and also on the properties of the inverse matrix $Q_{l l'}$.

\begin{figure}
\includegraphics[width=0.45\textwidth]{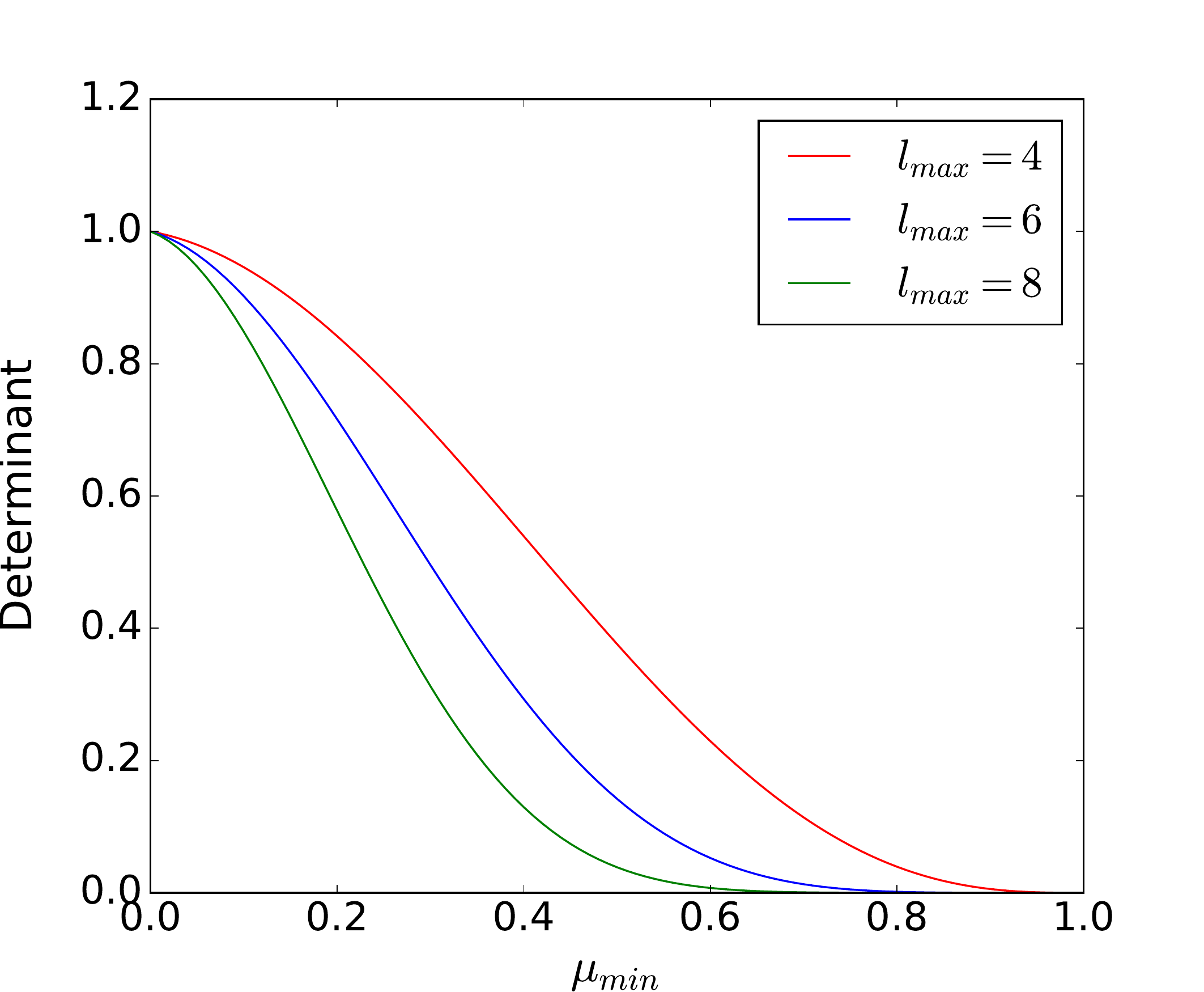}
\caption{The determinant of the bias matrix $q_{l l'}$ as a function of $\mu_{\rm min}$ for three different values of $l_{\rm max}$.}
\label{fig:qmat_det} 
\end{figure}

We have checked with our simulations and also using some analytical toy models whether the inversion given in \eqn{eq:Delta_sq_l_Delta_sq_l_win} is possible. We found that, for the range of redshifts considered in this paper, the series does not converge even for the optimistic case of $\mu_{\rm min} = 0.5$, and the recovered values of $\Delta^2_0(k)$ are typically off from the true value by at least $50 \%$. Since the convergence depends on the invertibility of the bias matrix $q_{l l'}$, we investigate this further by plotting its determinant as a function of $\mu_{\rm min}$. The results are shown in \fig{fig:qmat_det} where the determinant is plotted for three values of $l_{\rm max}$ (the maximum value of $l$ considered for calculating the determinant). As expected, the value of the determinant decreases with increasing $\mu_{\rm min}$, thus making the matrix less and less invertible. We also see that, for the same $\mu_{\rm min}$, the value of the determinant is smaller for larger $l_{\rm max}$. Thus including more terms in the series too makes the matrix less invertible, thus not allowing for the results to converge. We have checked and found that one can obtain reasonably acceptable results (say, within 10\% of the true value) for $\mu_{\rm min} \lesssim 0.2$ which corresponds to the value of the determinant $\gtrsim 0.8$. 

\begin{figure*}
\includegraphics[width=0.9\textwidth]{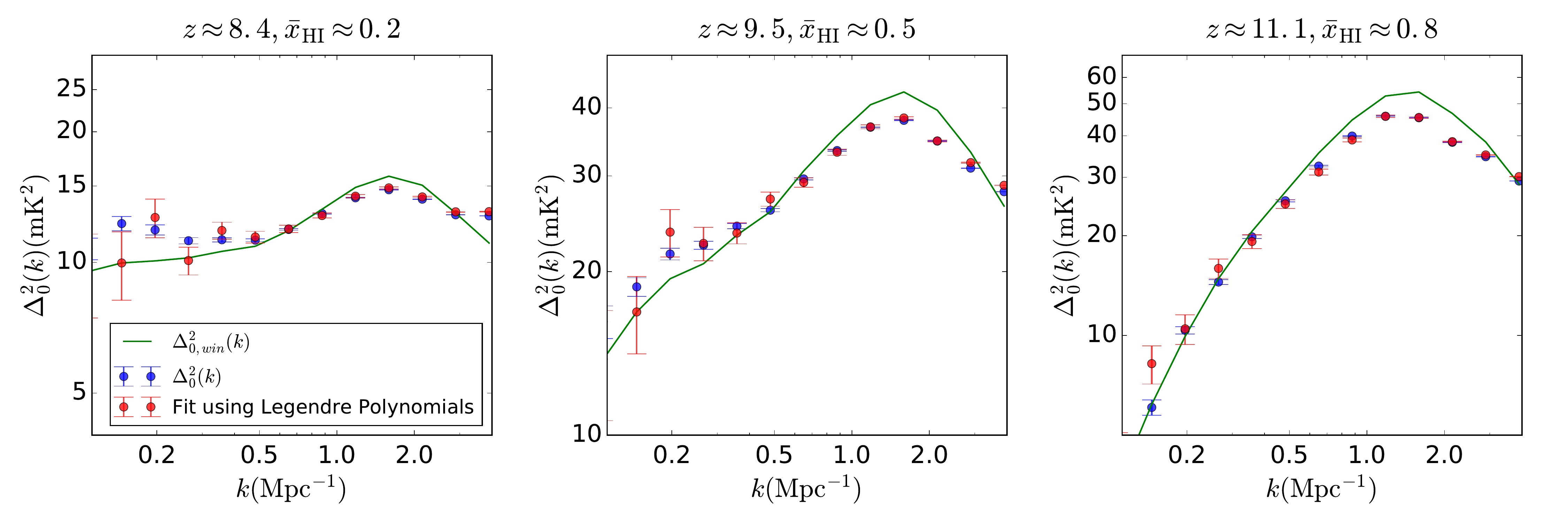}
\caption{The monopole $\Delta_0^2(k)$ computed for the case $\mu_{\rm min} = 0.5$ through fitting with Legendre polynomials in which the first three even multipoles were assumed to be nonzero. The redshifts and neutral hydrogen fractions for the three panels are same as earlier plots. The results of the fits are shown by red points with error-bars, where the errors are assumed to arise only from the fitting procedure. For reference, the monopole computed over the full $k$-space (blue points) and in the window (green lines) are also plotted.}
\label{fig:legfit_mumin0p5} 
\end{figure*}

As an alternative, one can also seek to fit the first three original Legendre Polynomials ${\cal P}_l(\mu)$ directly to the data in the window $\mu_{\rm min} \leq \mu \leq 1$ and thus obtain the true multipoles of the power spectrum. The results for the spherically averaged power spectrum $\Delta_0^2(k)$ for $\mu_{\rm min} = 0.5$ are shown in \fig{fig:legfit_mumin0p5}. The true power spectrum $\Delta_0^2(k)$ is denoted by blue points, while the red points are the ones obtained using the fitting procedure. The red error-bars correspond to the errors arising only from the fitting. We can see that the match between the fitted values and the true ones are quite good at smaller scales $k \gtrsim 0.3$ Mpc$^{-1}$. At larger scales, however, the values obtained from the fitting are different from the true ones. In addition, the fitting errors are also larger. The reason for this is that we have access to less number of $\mu$-modes at these scales for the boxes we are using. The disagreement between the fitted and the true values at large-scales could be an artefact of the box sizes employed, and perhaps will decrease for larger boxes. We plan to investigate such issues with simulations of higher dynamic range in a different project.

We have attempted to recover the higher moments (i.e., the quadrupole and the hexadecapole) too using the Legendre polynomial fits. The conclusions are broadly similar to what has been discussed above, i.e., the procedure works reasonably well at smaller scales but fails at large scales. One additional complication in the fitting procedure is that the errors in the different multipoles get correlated. This is because the Legendre Polynomials ${\cal P}_l(\mu)$ do not form orthogonal basis over the region of interest i.e. $\mu_{\rm min} \le \mu \le 1$. One needs to take this point into account while recovering the power spectra from the reionization window. Not surprisingly, the results remain identical if we use polynomial fitting instead of the Legendre polynomials.

We have also tried the same procedure for $\mu_{\rm min} = 0.95$. In this case, the number of $\mu$-modes available decrease significantly and the recovery is quite poor even for smaller scales. Hence we do not show the results in the paper.

\section{Conclusions}

The presence of various astrophysical foregrounds poses a severe challenge in detecting the 21~cm brightness temperature fluctuations from the epoch of reionization. Since the foregrounds are confined mostly to a wedge-shaped regions in the $k_{\perp} - k_{\parallel}$ space \citep{2010ApJ...724..526D,2012ApJ...745..176V,2012ApJ...752..137M,2012ApJ...757..101T,2012ApJ...756..165P,2013ApJ...768L..36P,2013ApJ...770..156H,2014PhRvD..90b3018L,2014PhRvD..90b3019L,2015ApJ...804...14T}, it has been proposed that the cosmological 21~cm power spectrum be measured using Fourier modes outside this region (known as the reionization window). Assuming that all the observational systematics are properly accounted for in the window and the reionization power spectrum is indeed measured, it still can lead to a bias in the computation of the spherically averaged power spectrum \citep{2016MNRAS.456...66J}. The reason for this wedge bias is that the presence of the peculiar velocity effects along the line of sight makes the underlying power spectrum anisotropic, hence the effect of the missing modes must be included in the analysis.

In this work, we have outlined a method which is useful for comparing theoretical model predictions (e.g., the power spectrum) with observational data when the foregrounds are dealt with by the avoidance technique. The method is based on the so-called clustering wedges \citep{2013MNRAS.435...64K} where one uses the angular moments to estimate the power spectrum in the reionization window. The basic idea behind our method is based on the fact that the spherically averaged power spectrum in the window picks up contribution from the true higher order multipoles. In order to keep the accounting simple, we have expanded the power spectrum in the reionization window using the basis of the \emph{shifted} Legendre polynomials. This basis has the advantage that it keeps the isolation of anisotropies intact in the window, and can be used for naturally extending the method to estimate the higher order multipoles in the window.

For estimating the power spectra using our method, in addition to the theoretical model predictions, we need to supply the value of $\mu_{\rm min}$ which measures the extent of the wedge in the Fourier space and compute the bias matrix $q_{l l'}$ defined in \eqn{eq:matrix_q}. Once these quantities are known, the clustering wedges in the window can be calculated by summing over all the significant multipoles, i.e., using the series in \eqn{eq:Delta_sq_l_win_Delta_sq_l}.

The main issue is to understand the number of terms to be retained in the series summation in \eqn{eq:Delta_sq_l_win_Delta_sq_l}. In the quasi-linear approximation, the series naturally terminates at the hexadecapole term ($l' = 4$). However, the presence of non-linearities in the HI fluctuation field $\delta_{\rm HI}$ can lead to higher order terms in the series. We investigate this issue using the semi-numerical simulations of reionization \citep{2015MNRAS.447.1806G,2015MNRAS.453.3143G} which incorporate the non-linearities expected in the model. We find that for values of $\mu_{\rm min} \lesssim 0.8$, terminating the series at $l' = 4$ provides good agreement between the clustering wedge and the true power spectrum in the window. This in turn implies that the method based on clustering wedges will eliminate any bias while interpreting the power spectrum measurements in the window.

One of the effects ignored in this study is that of the instrumental noise arising from the system temperature of the radio telescope. The presence of the noise would lead to larger errorbars than what has been assumed for the signal, however, we do not expect it to affect the wedge bias in any significant way. Also, we have ignored other line of sight effects like the light cone \citep{2006MNRAS.372L..43B,2012MNRAS.424.1877D,2014MNRAS.442.1491D,2014ApJ...789...31L,2014MNRAS.439.1615Z,2015MNRAS.453.3143G,2017arXiv170609449M}, which may have some impact on modelling the signal in presence of the wedge.  Also note that the cosmological signal may also get affected along with the foregrounds when one attempts foreground avoidance techniques. For example, to localize the foregrounds to the wedge region, one usually has to apply a window function which, to a certain extent, can also have the effect of degrading the cosmological power spectrum. We did not consider such subtle $k$-space blurring effects as they are beyond the scope of this paper.
\par As it is in any kind of experiments, one eventually aims to carry out a model comparison and obtain constraints on model parameters using the reionization 21~cm power spectrum \citep{2015MNRAS.449.4246G,2016MNRAS.455.4295G,2017MNRAS.468..122H,2017arXiv170503471G,2017MNRAS.468.3869S,2017arXiv170800011S}. It is expected that the wedge bias in the power spectrum will lead to bias in the parameter values. The clustering wedges discussed in this work are expected to bias-free and thus can be useful to constrain parameters in presence of the wedge. This would undoubtedly affect the statistical errors on the parameters as the measurements are based to fewer number of modes. We plan to study and quantify these effects in the future.

\section*{Acknowledgements}

The authors would like to thank Aseem Paranjape for discussions on the subject and Suman Majumdar for comments on the paper. We also thank the anonymous referee for constructive comments which helped improve the discussions in the paper. The simulations used in the paper were performed on the IBM cluster hosted by the National Centre for Radio Astrophysics, Pune, India.

\bibliographystyle{mnras}
\bibliography{main}

\bsp    
\label{lastpage}
\end{document}